\documentclass{ws-ijmpd}
\usepackage[super,compress]{cite}
\usepackage{url}
\usepackage{epstopdf}

\begin{document}

\markboth{R. C. G. Landim}
{Holographic dark energy from minimal supergravity}

%
\catchline{}{}{}{}{}
%

\title{Holographic dark energy from minimal supergravity}

\author{Ricardo C. G. Landim}

\address{Instituto de F\'isica, Universidade de S\~ao Paulo\\
 Caixa Postal 66318,  05314-970 S\~ao Paulo, S\~ao Paulo, Brazil\\
rlandim@if.usp.br}

\maketitle

\begin{history}
\received{Day Month Year}
\revised{Day Month Year}
\end{history}

\begin{abstract}
We embed models of holographic dark energy coupled to dark matter in minimal supergravity plus matter, with one chiral superfield. We analyze two cases. The first one has the Hubble radius as the infrared cutoff and the interaction between the two fluids is proportional to the energy density of the dark energy. The second case has the future event horizon as infrared cutoff while the interaction is proportional to the energy density of  both components of the dark sector.
\end{abstract}


\ccode{PACS numbers: 95.36.+x}

\section{Introduction}

Sixty eight percent of our universe\cite{Planck2013cosmological} consists of a still mysterious component called ``dark energy'' (DE), which is believed to be responsible for the  present acceleration of the universe\cite{reiss1998, perlmutter1999}. Among a wide range of alternatives for the dark energy  (see Ref.~\refcite{copeland2006dynamics} for review),  which includes the cosmological constant, scalar or vector fields, modifications of gravity and different kinds of cosmological fluids, the usage of a canonical scalar field, called ``quintessence'', is a viable and natural candidate\cite{peebles1988,ratra1988,Frieman1992,Frieman1995,Caldwell:1997ii}. Another striking attempt  to explain the acceleration comes from holography. The holographic principle states that the degrees of freedom of a physical system scales with its boundary area  rather than its volume\cite{'tHooft:1993gx,Susskind:1994vu}. Cohen and collaborators\cite{Cohen:1998zx} suggested that the dark energy should obey this principle, thus its energy density  has an upper limit and the fine-tuning problem for the cosmological constant is eliminated. Refs.~\refcite{Hsu:2004ri,Li:2004rb} followed the previous ideas regarding holography and argued that the holographic dark energy (HDE) has an energy density given by  $\rho_D=3c^2/L^2$ ($M_{p}^{-2}\equiv 8\pi G=1$), where $c$ is a constant and $L$ is an infrared (IR) cutoff. The first choice for $L$ was  the Hubble radius $H^{-1}$, however it led to an equation of state that describes dust\cite{Hsu:2004ri}. The correct equation of state for dark energy was obtained by \cite{Li:2004rb}, when he chose the future event horizon as the IR cutoff. 
 The problem with the Hubble radius as the IR cutoff could be avoided assuming the interaction between dark energy and dark matter (DM)\cite{Pavon:2005yx}. Such interaction was first proposed in the context of quintessence\cite{Wetterich:1994bg,Amendola:1999er}, and since the energy  densities of the DE and DM are comparable, the interaction can alleviate the coincidence problem\cite{Zimdahl:2001ar,Chimento:2003iea}. Taking the coupling into account, the HDE with Hubble radius as IR cutoff could lead to an accelerated expansion of the universe and also solve the coincidence problem\cite{Pavon:2005yx}. Li's proposal could also be generalized assuming the interaction between the two components of the dark sector\cite{Wang:2005jx,Wang:2005pk,Wang:2005ph,Wang:2007ak}. 
 
   From the point of view of theoretical physics, it would be interesting find out a model of dark energy from first principles. Since supergravity is  the low-energy limit of the superstring theory, it is natural to investigate if it can provide a model that describes the accelerated expansion of the universe. The simplest supergravity case is with one supersymmetry ($\mathcal{N}=1$)\footnote{ Extended supergravities can be applied to cosmology as well, see for instance \cite{kallosh2002supergravity,kallosh2002gauged}. They are closer to string theory, however minimal supergravity can work as an effective theory or an approximation.} and in this framework  Refs.~\refcite{Brax1999,Copeland2000,Landim:2015upa} presented some models that try to describe dark energy through quintessence.  Due to the prominent role of the AdS/CFT correspondence\cite{Maldacena:1997re} to relate both supergravity and holography concepts, it is natural to ask if there is any connection between supergravity (even for $\mathcal{N}=1$) and HDE. As showed in Ref.~\refcite{Sheykhi:2011cn}, it is possible to establish a connection between HDE and different kind of scalar fields, such as quintessence, tachyon and K-essence, through reconstructed scalar potentials. 
   
    In this paper we show that the HDE in interaction with DM can be embedded in minimal supergravity plus matter with a single chiral superfield. The interaction between HDE and DM is taken into account because they are generalizations of some uncoupled cases, namely, the models presented in Ref.~\refcite{Hsu:2004ri,Li:2004rb}. We use Planck units ($\hbar=c=8\pi G=1$) throughout the text.

  We assume that the dark energy is coupled with dark matter, in such a way that total energy-momentum is still conserved. In the flat Friedmann-Robertson-Walker background with a scale factor $a$, the continuity equations for both components are

\begin{equation}\label{contide}
\dot{\rho_D}+3H(1+w_D)\rho_D=-Q, \quad \dot{\rho_M}+3H\rho_M=Q,
\end{equation}

\noindent  where $H=\dot{a}/a$ is the Hubble rate,  $Q$ is the coupling and the dot is a derivative with respect to the cosmic time $t$. The index $M$ stands for dark matter. The case $Q>0$ corresponds to dark energy transformation into dark matter, while $Q<0$ is the transformation in the opposite direction. In principle, the coupling  can depend on several variables $Q=Q(\rho_m,\rho_\phi, H,\dots)$, but assuming it is a small quantity, the interaction term can be written as a Taylor expansion, giving rise to three main kernels: (i) $Q\propto H\rho_M$, (ii)   $Q\propto H\rho_D$ and (iii)  $Q\propto H(\rho_M+\rho_D)$. In the context of HDE, the second kernel was used in Ref.~\refcite{Pavon:2005yx}), while the third one is found in Ref.~\refcite{Wang:2005jx,Wang:2007ak}. Both kernels generalize  previous results\cite{Hsu:2004ri,Li:2004rb}, respectively, thus the correspondent models are good choices to analyze whether they can be embedded in minimal supergravity or not.

\section{HDE with Hubble radius as IR cutoff}

 First we will consider the size $L$  as the Hubble radius $H^{-1}$, thus the energy density for the dark energy becomes
 
  \begin{equation}\rho_D=3c^2H^2, \label{eqH}
   \end{equation} 
  
  \noindent which describes a pressureless fluid\cite{Hsu:2004ri} in the absence of interaction with dark matter. When one considers such interaction\cite{Pavon:2005yx}, the fluid  has an equation of state that can describe the dark energy. We consider the interaction $Q=3b^2H\rho_D$, where $b$ is a constant which measures the strength of the interaction. The equation of state for dark energy with this interaction is written in terms of the ratio of the energy densities $r\equiv \rho_M/\rho_D=(1-c^2)/c^2$

   \begin{equation}w_D=-\left(1+\frac{1}{r}\right)b^2=-\frac{b^2}{1-c^2}.
   \label{eqPavon}
   \end{equation}

In order to reconstruct a holographic scalar field model we should relate (\ref{eqPavon}) with the scalar field $\phi$. Using the energy density and the pressure for the scalar field $\rho_\phi=\dot{\phi}^2/2+V(\phi)$ and $p_\phi=\dot{\phi}^2/2-V(\phi)$, we have

   \begin{equation}\dot{\phi}^2=(1+w_\phi)\rho_\phi, \quad V(\phi)=(1-w_\phi)\frac{\rho_\phi}{2}.
   \label{scalarV}
   \end{equation}

Using (\ref{eqH}) and (\ref{eqPavon}) into (\ref{scalarV}) we have the potential\cite{Sheykhi:2011cn}

 \begin{equation}
 V(\phi)=B^{-2}e^{-\sqrt{2}B\phi},
   \label{V1}
   \end{equation}
   
   \noindent where $B=\frac{3k}{2\sqrt{2}c}\left(3-\frac{3b^2}{1-c^2}\right)^{-1/2}$ and $k=1-b^2c^2/(1-c^2)$.

On the other hand, the scalar potential in minimal supergravity with no D-terms is given by 
  
\begin{equation}V= e^{K}\left( K^{\Phi\bar{\Phi}}\left|W_\Phi+K_\Phi W\right|^2-3|W|^2\right),\label{potential}
   \end{equation} 
   
    \noindent for a single complex scalar field. The potential above  depends on a real function $K\equiv K(\Phi,\bar{\Phi})$, called K\"ahler potential, and a holomorphic function $W\equiv W(\Phi)$, the superpotential. $K_{\Phi\bar{\Phi}}\equiv \frac{\partial^2 K}{\partial \Phi\partial\bar{\Phi}}$ is the K\"ahler metric, $W_\Phi\equiv\frac{\partial W}{\partial \Phi}$ and $K_\Phi\equiv\frac{\partial K}{\partial \Phi}$. We will use the same choices of Ref.~\refcite{Copeland2000} for $K$ and $W$, namely, the string-inspired K\"ahler potential $K=-\ln(\Phi+\bar{\Phi})$, which is present at the tree level for axion-dilaton field in string theory, and the superpotential $W=\Lambda^2\Phi^{-\alpha}$, where $\Lambda$ is a constant. With these choices, with the imaginary part of the scalar field stabilized at zero $\langle$Im $\Phi\rangle=0$ and the field redefinition Re $\Phi=e^{\sqrt{2}\phi}$, we get the scalar potential expressed in terms of the canonical normalized field $\phi$ 
   
 \begin{equation}V(\phi)= \frac{\Lambda^4}{2}(\beta^2-3)e^{-\sqrt{2}\beta\phi},\label{potentialsugra}
   \end{equation} 
   
   \noindent where  $\beta=2\alpha+1$. Comparing (\ref{V1}) and (\ref{potentialsugra}) we have $B=\beta$ and $\Lambda^2=\sqrt{2}/(\beta\sqrt{\beta^2-3})$, with $\beta^2\geq 3$. Thus, the two parameters $b$ and $c$ determine the exponent $\alpha$ of the superpotential and the parameter $\Lambda$. We notice in this case that $\Lambda$ is not an independent parameter, but it depends on $b$ and $c$ as well. Even if we would have a way to know what is the value of $\alpha$ or rather, $\beta$, we could not know the specific values of $b$ and $c$. The opposite direction is favored, because once one finds out observationally $b$ and $c$, the HDE can be embedded in a specific supergravity model. If there is no interaction ($b=0$) $\beta=\sqrt{3}/(2c\sqrt{2})$, but $w_D=0$.

   The kernel used so far simplifies $w_D$ in such a way that (\ref{V1}) was written with no need of any approximation. When one considers the other possibilities for the interaction $Q$ [(i) or (iii)], the scalar potential cannot be written as easy as it was in (\ref{V1}). We illustrate this possibility now, but with other IR cutoff.  
   
  \section{HDE with future event horizon as IR cutoff}
   We will analyze another possibility of HDE with the kernel $Q=3b^2H(\rho_M+\rho_D)$ and with the future event horizon $R_E=a\int_t^\infty{dt/a}=c\sqrt{1+r}/H$ as the IR cutoff.  The choice of $L$ is done because when $b=0$ the original Li's model of HDE\cite{Li:2004rb} is recovered. The energy density for the HDE is 
   
    \begin{equation}\label{rho2}
   \rho_D=\frac{3H^2}{1+r}
   \end{equation}
   
\noindent and the equation of state for HDE becomes
   
     \begin{equation}
     w_D=-\frac{1}{3}\left(1+\frac{2\sqrt{\Omega_D}}{c}+\frac{3b^2}{\Omega_D}\right),
   \label{eqabdalla}
   \end{equation}
   
  \noindent where $\Omega_D\equiv\rho_D/(\rho_D+\rho_M)=(1+r)^{-1}$ is the density parameter for DE. Using (\ref{rho2}) and (\ref{eqabdalla}) in (\ref{scalarV}) we have

        \begin{equation}
    \dot{\phi}^2=2H^2\Omega_D\left(1-\frac{\sqrt{\Omega_D}}{c}-\frac{3b^2}{2\Omega_D}\right),
   \label{phidot}
   \end{equation}
   
      \begin{equation}
    V(\phi)=H^2\Omega_D\left(2+\frac{3\sqrt{\Omega_D}}{c}+\frac{9b^2}{2\Omega_D}\right).
   \label{V}
   \end{equation}
   
 \noindent From the equations above we see that we have to know the evolution of $\Omega_D$ in order to fully determine  $V(\phi)$ as a function of $\phi$. The evolution equation  for $\Omega_D$ was found in Ref.~\refcite{Wang:2005jx} and it is
 
      \begin{equation}
   \frac{\Omega_D'}{\Omega_D^2}=(1-\Omega_D)\left(\frac{1}{\Omega_D}+\frac{2}{c\sqrt{\Omega_D}}-\frac{3b^2}{\Omega_D(1-\Omega_D)}\right),
   \label{omegaevolut}
   \end{equation}

   \noindent where the prime is the derivative with respect to $\ln a$. To solve this differential equation we let $y=1/\sqrt{\Omega_D}$, thus the equation (\ref{omegaevolut}) becomes

    \begin{equation}
   y^2y'=(1-y^2)\left(\frac{1}{c}+\frac{y}{2}+\frac{3b^2y^3}{2(1-y^2)}\right).
   \label{yevolut}
   \end{equation}

 In order to have an analytic solution of (\ref{yevolut}) we will investigate the asymptotic behavior of $\Omega_D$, for small and large $a$. For very small $a$ we have $\Omega_D\rightarrow 0$ and $y\rightarrow \infty$, therefore (\ref{yevolut}) is approximately

    \begin{equation}
   y'\approx(1-y^2)\left(\frac{3b^2}{2}-\frac{1}{2}\right)y,
   \label{yevolut2}
   \end{equation}
  
  \noindent which leads to the solution $\Omega_D=\Omega_0 a^{-(3b^2-1)}\approx \Omega_0 a $, provided that $b$ should be small\cite{Wang:2005jx}. Since  $\Omega_D$ scales with $a$, the contribution of the dark energy in the early universe is negligible. We also see from (\ref{phidot}) that the term between brackets should be positive and small, then $ \dot{\phi}\propto H\sqrt{2\Omega_D} $ is small. Therefore, we have $\phi(a)\sim a^{1/2}$. Due to the small contribution of  $\Omega_D$, we neglect it at this limit and we will focus on large $a$, where HDE is dominant. 
  
  For large $a$ we have $\Omega_D\rightarrow 1$, and $y\rightarrow 1$. Thus, the last term in the equation (\ref{yevolut}) is the dominant one, so $y'\approx\frac{3b^2}{2}y$, and $\Omega_D=\Omega_0a^{-3b^2}$, where $\Omega_0$ is the value of $\Omega_D$ at $a_0=1$. Since $\Omega_D=\Omega_0a^{-3b^2}=\Omega_0a^{-3(1+w_D)}$, we have $1+w_D=b^2$ which is very small if we consider the present value of $w_D$ given by Planck\cite{Planck2013cosmological}. Thus we neglect the last term in (\ref{phidot}) and we assume the dark energy dominance. We have  $\Omega_D=\Omega_0\approx 1$ at large $a$ and (\ref{phidot}) yields

   \begin{equation}
    \dot{\phi}^2\approx 2H^2\Omega_D\left(1-\frac{1}{c}\right).
   \label{phidot2}
   \end{equation}

  \noindent The equation above implies that $w_D\approx -\frac{1}{3}(1+2/c)$, which is the Li's proposal\cite{Li:2004rb}. Equation (\ref{phidot2}) has the solution

     \begin{equation}
   \phi(a)=\left(1-\frac{1}{c}\right)^{1/2}\sqrt{2\Omega_D}\ln a.
   \label{phidota}
   \end{equation}
   
   The second Friedmann equation at large $a$ leads to  $H=\frac{3c}{(c-1)t}$ and $a=t^{3/(c-1)}$. The scalar potential (\ref{V}) becomes

    \begin{equation}
V(\phi)\approx\left(2+\frac{3}{c}\right)H^2=\left(2+\frac{1}{c}\right)\frac{9c^2}{(c-1)^2}e^{-\frac{1}{3}\left(\frac{c-1}{c}\right)^{1/2}\phi}.\label{V2}
   \end{equation}
   
   Similarly to before, comparing (\ref{V2}) with (\ref{potentialsugra}) we see that both $\beta$ and $\Lambda$ are determined by the constant $c$. Equation (\ref{V2}) was deduced for a dark-energy-dominated universe, in such a way that the interaction with dark matter is absent, as it should be in this limit.

   A more realistic scenario could not be found analytically, in opposite to the previous case, where the scalar potential was written with no approximations.  Different kernels for the coupling $Q$ may be tried, although the main features of the method were expressed in these two cases. The other alternatives are similar.

  \section{Conclusions}
 In this paper we embedded two models of HDE in the minimal supergravity with one single chiral superfield. In the first model we used the Hubble radius as the IR cutoff and an interaction proportional to $\rho_D$, while in the second one we used the future event horizon as IR cutoff and the kernel proportional to $\rho_D+\rho_M$. In both cases the free parameters  of our superpotential could be expressed in terms of the constants $c$ and $b$, depending on the model. The second case is embedded only for a dark-energy-dominated universe, while the first one is more general. There are other ways to embed HDE in minimal supergravity, as for instance the choices of K\"ahler potential and superpotential made in Ref.~\refcite{Nastase:2015pua} for the inflationary scenario. However, the main results are the same, that is, a way to relate HDE and supergravity. Due to the nature of the holographic principle and recalling that extended supergravity is the low-energy limit of string theory, the relation presented here has perhaps a deeper meaning, when one takes a quantum gravity theory into account.

\section*{Acknowledgments}
I thank Elcio Abdalla for comments. This work is supported by FAPESP Grant No. 2013/10242-1.

\bibliographystyle{ws-ijmpd}
\bibliography{trab1}

\providecommand{\noopsort}[1]{}\providecommand{\singleletter}[1]{#1}%
\begin{thebibliography}{10}

\bibitem{Planck2013cosmological}
 Planck Collaboration (P.~A.~R. Ade {\em et~al.}), {\em Astron.Astrophys.} {\bf
  571}  (2014)   A16, \href{http://arxiv.org/abs/1303.5076}{{\ttfamily
  arXiv:1303.5076 [astro-ph.CO]}}.

\bibitem{reiss1998}
 Supernova Search Team Collaboration (A.~G. Riess {\em et~al.}), {\em
  Astron.J.} {\bf 116}  (1998) 1009,
  \href{http://arxiv.org/abs/astro-ph/9805201}{{\ttfamily
  arXiv:astro-ph/9805201 [astro-ph]}}.

\bibitem{perlmutter1999}
 Supernova Cosmology Project Collaboration (S.~Perlmutter {\em et~al.}), {\em
  Astrophys.J.} {\bf 517}  (1999) 565,
  \href{http://arxiv.org/abs/astro-ph/9812133}{{\ttfamily
  arXiv:astro-ph/9812133 [astro-ph]}}.

\bibitem{copeland2006dynamics}
E.~J. Copeland, M.~Sami and S.~Tsujikawa, {\em Int. J. Mod. Phys.} {\bf D15}
  (2006) 1753, \href{http://arxiv.org/abs/hep-th/0603057}{{\ttfamily
  arXiv:hep-th/0603057 [hep-th]}}.

\bibitem{peebles1988}
P.~Peebles and B.~Ratra, {\em Astrophys.J.} {\bf 325}  (1988)   L17.

\bibitem{ratra1988}
B.~Ratra and P.~Peebles, {\em Phys.Rev.} {\bf D37}  (1988)   3406.

\bibitem{Frieman1992}
J.~A. Frieman, C.~T. Hill and R.~Watkins, {\em Phys.Rev.} {\bf D46}  (1992)
  1226.

\bibitem{Frieman1995}
J.~Frieman, C.~Hill, A.~Stebbins and I.~Waga, {\em Phys. Rev. Lett.} {\bf 75}
  (1995)   2077.

\bibitem{Caldwell:1997ii}
R.~Caldwell, R.~Dave and P.~Steinhardt, {\em Phys. Rev. Lett.} {\bf 80}  (1998)
    1582.

\bibitem{'tHooft:1993gx}
G.~'t~Hooft, { {Dimensional reduction in quantum gravity}}, in {\em {Salamfest
  1993:0284-296}\/},  (1993), pp. 0284--296.
\newblock \href{http://arxiv.org/abs/gr-qc/9310026}{{\ttfamily
  arXiv:gr-qc/9310026 [gr-qc]}}.

\bibitem{Susskind:1994vu}
L.~Susskind, {\em J. Math. Phys.} {\bf 36}  (1995) 6377,
  \href{http://arxiv.org/abs/hep-th/9409089}{{\ttfamily arXiv:hep-th/9409089
  [hep-th]}}.

\bibitem{Cohen:1998zx}
A.~G. Cohen, D.~B. Kaplan and A.~E. Nelson, {\em Phys. Rev. Lett.} {\bf 82}
  (1999) 4971, \href{http://arxiv.org/abs/hep-th/9803132}{{\ttfamily
  arXiv:hep-th/9803132 [hep-th]}}.

\bibitem{Hsu:2004ri}
S.~D.~H. Hsu, {\em Phys. Lett.} {\bf B594}  (2004) 13,
  \href{http://arxiv.org/abs/hep-th/0403052}{{\ttfamily arXiv:hep-th/0403052
  [hep-th]}}.

\bibitem{Li:2004rb}
M.~Li, {\em Phys. Lett.} {\bf B603}  (2004)  ~1,
  \href{http://arxiv.org/abs/hep-th/0403127}{{\ttfamily arXiv:hep-th/0403127
  [hep-th]}}.

\bibitem{Pavon:2005yx}
D.~Pavon and W.~Zimdahl, {\em Phys. Lett.} {\bf B628}  (2005) 206,
  \href{http://arxiv.org/abs/gr-qc/0505020}{{\ttfamily arXiv:gr-qc/0505020
  [gr-qc]}}.

\bibitem{Wetterich:1994bg}
C.~Wetterich, {\em Astron.Astrophys.} {\bf 301}  (1995) 321,
  \href{http://arxiv.org/abs/hep-th/9408025}{{\ttfamily arXiv:hep-th/9408025
  [hep-th]}}.

\bibitem{Amendola:1999er}
L.~Amendola, {\em Phys.Rev.} {\bf D62}  (2000)   043511,
  \href{http://arxiv.org/abs/astro-ph/9908023}{{\ttfamily
  arXiv:astro-ph/9908023 [astro-ph]}}.

\bibitem{Zimdahl:2001ar}
W.~Zimdahl and D.~Pavon, {\em Phys.Lett.} {\bf B521}  (2001) 133,
  \href{http://arxiv.org/abs/astro-ph/0105479}{{\ttfamily
  arXiv:astro-ph/0105479 [astro-ph]}}.

\bibitem{Chimento:2003iea}
L.~Chimento, A.~Jakubi, D.~Pavon and W.~Zimdahl, {\em Phys.Rev.} {\bf D67}
  (2003)   083513, \href{http://arxiv.org/abs/astro-ph/0303145}{{\ttfamily
  arXiv:astro-ph/0303145 [astro-ph]}}.

\bibitem{Wang:2005jx}
B.~Wang, Y.-G. Gong and E.~Abdalla, {\em Phys. Lett.} {\bf B624}  (2005) 141,
  \href{http://arxiv.org/abs/hep-th/0506069}{{\ttfamily arXiv:hep-th/0506069
  [hep-th]}}.

\bibitem{Wang:2005pk}
B.~Wang, Y.~Gong and E.~Abdalla, {\em Phys. Rev.} {\bf D74}  (2006)   083520,
  \href{http://arxiv.org/abs/gr-qc/0511051}{{\ttfamily arXiv:gr-qc/0511051
  [gr-qc]}}.

\bibitem{Wang:2005ph}
B.~Wang, C.-Y. Lin and E.~Abdalla, {\em Phys. Lett.} {\bf B637}  (2006) 357,
  \href{http://arxiv.org/abs/hep-th/0509107}{{\ttfamily arXiv:hep-th/0509107
  [hep-th]}}.

\bibitem{Wang:2007ak}
B.~Wang, C.-Y. Lin, D.~Pavon and E.~Abdalla, {\em Phys. Lett.} {\bf B662}
  (2008) 1, \href{http://arxiv.org/abs/0711.2214}{{\ttfamily arXiv:0711.2214
  [hep-th]}}.

\bibitem{kallosh2002supergravity}
R.~Kallosh, A.~D. Linde, S.~Prokushkin and M.~Shmakova, {\em Phys. Rev.} {\bf
  D66}  (2002)   123503, \href{http://arxiv.org/abs/hep-th/0208156}{{\ttfamily
  arXiv:hep-th/0208156 [hep-th]}}.

\bibitem{kallosh2002gauged}
R.~Kallosh, A.~D. Linde, S.~Prokushkin and M.~Shmakova, {\em Phys. Rev.} {\bf
  D65}  (2002)   105016, \href{http://arxiv.org/abs/hep-th/0110089}{{\ttfamily
  arXiv:hep-th/0110089 [hep-th]}}.

\bibitem{Brax1999}
P.~Brax and J.~Martin, {\em Phys. Lett.} {\bf B468}  (1999) 40,
  \href{http://arxiv.org/abs/astro-ph/9905040}{{\ttfamily
  arXiv:astro-ph/9905040 [astro-ph]}}.

\bibitem{Copeland2000}
E.~J. Copeland, N.~J. Nunes and F.~Rosati, {\em Phys. Rev.} {\bf D62}  (2000)
  123503, \href{http://arxiv.org/abs/hep-ph/0005222}{{\ttfamily
  arXiv:hep-ph/0005222 [hep-ph]}}.

\bibitem{Landim:2015upa}
R.~C.~G. Landim  (2015) \href{http://arxiv.org/abs/1509.04980}{{\ttfamily
  arXiv:1509.04980 [hep-th]}}.

\bibitem{Maldacena:1997re}
J.~M. Maldacena, {\em Int. J. Theor. Phys.} {\bf 38}  (1999) 1113,
  \href{http://arxiv.org/abs/hep-th/9711200}{{\ttfamily arXiv:hep-th/9711200
  [hep-th]}}, [Adv. Theor. Math. Phys.2,231(1998)].

\bibitem{Sheykhi:2011cn}
A.~Sheykhi, {\em Phys. Rev.} {\bf D84}  (2011)   107302,
  \href{http://arxiv.org/abs/1106.5697}{{\ttfamily arXiv:1106.5697
  [physics.gen-ph]}}.

\bibitem{Nastase:2015pua}
H.~Nastase, {\em Nucl. Phys.} {\bf B903}  (2016) 118,
  \href{http://arxiv.org/abs/1504.06497}{{\ttfamily arXiv:1504.06497
  [hep-th]}}.

\end{thebibliography}

\end{document}